\documentstyle[epsfig,preprint,aps]{revtex}
\begin{document}
\draft
\preprint{}

\title{Polarization effects in electroweak \\
       vector boson productions}
\vskip 0.5cm

\author{J. S. Shim}
\address{Department of Physics, Kyung Hee University,
         Seoul 130-701, Korea}

\author{S. W. Baek and H. S. Song}
\address{Center for Theoretical Physics and Department of
         Physics,\\
         Seoul National University, Seoul 151-742, Korea}

\maketitle

\begin{abstract}
The electroweak production processes of vector bosons,
$e^+ e^- \rightarrow Z \gamma$, $\gamma e \rightarrow Ze$,
and $\gamma e \rightarrow W \nu$ are considered
simultaneousely and in the Standard model the covariant
polarization density matrices of vector bosons for these
processes are
obtained explicitly from one of the processes, $\gamma e
\rightarrow Z e$. The effect of the photon
polarization as well as the vector meson polarization is
considered and some special cases are discussed.
\end{abstract}

\newpage


\section{Introduction}
\label{sec:intro}


The standard model (SM) provides an excellent description
of physics below the electroweak scale but we also believe
that new physics beyond the SM must exist at higher energy.
Three processes, $e^+ e^- \rightarrow Z \gamma$, $\gamma e
\rightarrow Z e$, and $\gamma e \rightarrow W \nu$ are the
simplest ones to check the SM.
The $e^+ e^- \rightarrow Z \gamma$ process can be
investigated experimentally at relatively low energy scale.
This process was also considered to search for the number of
neutrino generations \cite{EMa,GB} since it is shown
\cite{GB}
that the major contribution to the process $e^+ e^-
\rightarrow \nu \bar{\nu} \gamma$ at the low energy is the
$Z^0$ exchange, i.e., near the $Z^0$ peak the $e^+ e^-
\rightarrow Z^0 \gamma$ and the subsequent decay of $Z^0$
into $\nu \bar{\nu}$.
The processes $\gamma e \rightarrow Z e$ and $\gamma e
\rightarrow  W \nu$ might be possible experimetally by using
the backscattered laser beams at the future linear $e^+ e^-$
colliders.

These processes have been discussed in order to investigate
the effect of possible deviations from the SM through
the anomalous terms of vector meson couplings.
The anomalous $\gamma\gamma Z$ and $\gamma ZZ$ vertices
are considered in the process $e^+ e^- \rightarrow \gamma Z$
\cite{MG,DC} and also in the process $\gamma e \rightarrow
Z e$ \cite{FMR2,AD,SY1}.
In the process $\gamma e \rightarrow W \nu$, the anomalous
$\gamma WW$ terms are considered \cite{LF,AG,EY,MR}

While the $e^+ e^- \rightarrow Z \gamma$ process can be
described in the c.m. frame, the $\gamma e \rightarrow Z e$
and $\gamma e\rightarrow W \nu$ processes are not necessarily
in the c.m. frame since the energy of the Compton
backscattered photons is less than that of the initial
electron beam. Usually the helicity formalism in the c.m.
frame has been used.

The purpose of this paper is to consider the processes
$e^+ e^-\rightarrow Z \gamma$, $\gamma e \rightarrow Z e$,
and $\gamma e \rightarrow W \nu$ simultaneously in an
arbitrary frame.
The covariant density matrix formalism \cite{SY2} is used
to obtain the polarization of vector mesons.
In the SM, it is shown that these three processes are
interrelated and, therefore, starting
from the process $\gamma e \rightarrow Z e$, the result of
other two processes can be obtained,
even though the $\gamma e \rightarrow Z e$ and $\gamma e
\rightarrow W \nu$ processes are described by different
Feynman diagrams, i.e., in the $\gamma e \rightarrow Z e$
there are no boson coupling terms like $WW\gamma$ term in
the $\gamma e \rightarrow W \nu$.
The process $e^+ e^- \rightarrow Z \gamma$ considered in the
c.m. frame previously by us \cite{JC} can be reproduced
from a general and much simpler form obtained from the
$\gamma e \rightarrow Z e$ process.

The general consideration of these processes in the SM can be
useful in considering the new physics beyond the SM and it is
also useful to discuss the contribution of the processes to
the other final states like the $\gamma e \rightarrow \mu \nu
\bar{\nu}$ even in the SM.
If the anomalous vector coupling terms in these
processes are considered beyond the SM, the
situation is changed and the effect of additional terms
can be checked through the polarization of vector bosons.
The method discussed here can be used for the investigation
of possible deviations from the SM.

\newpage


\section{Transition Amplitude of Vector Boson Production}
\label{sec:tran}


In the SM, the transition amplitude of the $\gamma e
\rightarrow Z e$ process is described in the tree level
by the transition amplitude
\begin{eqnarray}
&&{\cal M}_{\gamma e\rightarrow Ze}=
\frac{e}{2}ig_Z
   \bar{u}(p_2,s_2)
\left[\not\!{\varepsilon^*_Z}
     \frac{(\not\!{k_1}\not\!{\varepsilon_\gamma}+2p_1\cdot
       \varepsilon_\gamma)} {k_1\cdot p_1}
 - \frac{(2p_2\cdot \varepsilon_\gamma
       -\not\!{\varepsilon_\gamma} \not\!{k_1})}
    {k_1\cdot p_2}
   \not\!{\varepsilon^*_Z}\right]
\nonumber\\
&&\hskip 5cm
  \times
  \left[\epsilon_L(1-\gamma_5) +\epsilon_R(1+\gamma_5)\right]
      u(p_1,s_1).\label{reze}
\end{eqnarray}
Here $k_1, p_1, k_2$, and $p_2$ are momenta of incoming photon,
incoming electron, outgoing $Z^0$, and outgoing electron,
respectively. Also $\varepsilon_\gamma ^\mu$ and
$\varepsilon^{\mu}_Z$ are the wave vectors of photon and $Z^0$,
respectively, and $g_Z$, $\epsilon_L$, and $\epsilon_R$ are
defined as
\begin{eqnarray}
&&g_Z =\left(\sqrt{2} G_F m_Z^2 \right)^{\frac{1}{2}},
\nonumber\\
&&\epsilon_L=-\frac{1}{2}+\sin^2\theta_W,\nonumber\\
&& \epsilon_R=\sin^2\theta_W.
\end{eqnarray}

The process $\gamma e \rightarrow W \nu$ is described by
the transition amplitude
\begin{eqnarray}
&&{\cal M}_{\gamma e\rightarrow W\nu_e}=
\frac{e}{2}ig_W
   \bar{u}(p_2,s_2)
\left\{\not\!{\varepsilon^*_W}
        \frac{(\not\!{k_1} \not\!{\varepsilon_\gamma}
           +2p_1\cdot \varepsilon_\gamma)}
           {k_1\cdot p_1}
\right. \nonumber\\
&&\hskip 2cm \left.
 + \frac{2}{k_1\cdot k_2}
\left[(\varepsilon_\gamma \cdot\varepsilon^*_W)\not\!{k_1}
  -(\varepsilon^*_W\cdot k_1)\not\!{\varepsilon_\gamma}
  -(\varepsilon_\gamma\cdot k_2)\not\!{\varepsilon^*_W}
      \right]
      \right\}(1 - \gamma_5) u(p_1,s_1),
\label{trew}
\end{eqnarray}
where
\begin{eqnarray}
&&g_W
=\left( 2^{-\frac{1}{2}} G_F m_W^2 \right)^{\frac{1}{2}}
      =\frac{g_Z \cos\theta_W}{\sqrt{2}}.
\end{eqnarray}

The transition amplitudes for the $\gamma e \rightarrow Z e$
and $\gamma e \rightarrow W \nu$ look quite different due to
their Feynman diagrams as shown as Figs. 1 and 2.
But the term in the bracket of Eq.~(\ref{trew}) can be
converted into the form in the bracket of
Eq.~(\ref{reze}) by using the relation,
\begin{eqnarray}
&&\left\{\not\!{\varepsilon^*_W}
        \frac{(\not\!{k_1} \not\!{\varepsilon_\gamma}
           +2p_1\cdot \varepsilon_\gamma)}
           {k_1\cdot p_1}
 + \frac{2}{k_1\cdot k_2}
\left[(\varepsilon_\gamma \cdot\varepsilon^*_W)\not\!{k_1}
  -(\varepsilon^*_W\cdot k_1)\not\!{\varepsilon_\gamma}
  -(\varepsilon_\gamma\cdot k_2)\not\!{\varepsilon^*_W}
      \right]
      \right\}\nonumber\\
&&=-\frac{k_1\cdot p_2}{k_1\cdot k_2}
    \left[\not\!{\varepsilon^*_W}
     \frac{(\not\!{k_1}\not\!{\varepsilon_\gamma}+2p_1\cdot
       \varepsilon_\gamma)} {k_1\cdot p_1}
 - \frac{(2p_2\cdot \varepsilon_\gamma
       -\not\!{\varepsilon_\gamma} \not\!{k_1})}
    {k_1\cdot p_2}
   \not\!{\varepsilon^*_W}\right].
\end{eqnarray}
Therefore, various results obtained in the
$\gamma e \rightarrow Z
e$ process can be used in the $\gamma e \rightarrow W \nu_e$
process after $\epsilon_R, m_Z$, and $\epsilon_L g_Z$ are
replaced
by $0, m_W$, and $-(k_1\cdot p_2/k_1\cdot k_2) g_W$
, respectively.

Since the transition amplitude of the $e^+ e^- \rightarrow
Z \gamma$ process is described by
\begin{eqnarray}
&&{\cal M}_{e^+ e^-  \rightarrow Z \gamma}=
\frac{e}{2}ig_Z
   \bar{v}(p_2,s_2)
\left[\not\!{\varepsilon^*_Z}
     \frac{(\not\!{k_1}\not\!{\varepsilon_\gamma^*}-2p_1\cdot
       \varepsilon_\gamma^*)} {k_1\cdot p_1}
 + \frac{(2p_2\cdot \varepsilon_\gamma^*
       -\not\!{\varepsilon_\gamma^*} \not\!{k_1})}
    {k_1\cdot p_2}
   \not\!{\varepsilon^*_Z}\right]
\nonumber\\
&&\hskip 5cm
  \times
  \left[\epsilon_L(1-\gamma_5) +\epsilon_R(1+\gamma_5)\right]
      u(p_1,s_1),\label{eeZr}
\end{eqnarray}
where $p_1, p_2, k_1$, and $k_2$ are momenta of positron
, electron, photon, and $Z^0$, respectively,
one can see that it can be obtained from Eq.~(\ref{reze})
just by replacing $k_1, p_2, \varepsilon_\gamma$, and
$\bar{u}(p_2, s_2)$ by $-k_1, -p_2$,
$\varepsilon^*_\gamma$, and $\bar{v}(p_2,s_2)$,
respectively.
The differential cross sections as well as polarization vectors
and tensors of vector
bosons can be obtained from those obtained in the process
$\gamma e \rightarrow Z e$.
The transition amplitudes become simplified if the helicity
formalism is used, but the covariant formalism is used
here in order to treat the $e^+ e^- \rightarrow Z \gamma$,
$\gamma e \rightarrow Z e$, and $\gamma e \rightarrow W \nu$
simultaneously.


\section{Polarization Effects}
\label{sec:pol}


Once the transition amplitude producing the vector mesons
is given, it is straightforward to obtain the polarization
vector and tensor of the spin-1 particle according to the
following method.

The wave vector $\varepsilon^\mu(p,\lambda)$ of a massive
spin-1 particle is expressed as \cite{SY2}
\begin{eqnarray}
\varepsilon^\mu(p,\lambda)=(1-|\lambda|)n_3^\mu
                 +\frac{1}{\sqrt{2}}|\lambda|
                  (-\lambda n_1^\mu -i n_2^\mu),
		  \label{Wvec}
\end{eqnarray}
where $p$ and $\lambda$ are its four-momentum and spin
components, $\pm 1$ or $0$, respectively, and the $n_i^\mu$
$(i=1,2,3)$
are the Lorentz boosts of an arbitrary orthogonal basis
($\hat{\theta}, \hat{\phi}, \hat{s}$) in three dimensional
space. Together with $n_0^\mu=p^\mu /m$, the $n_i^\mu$
form a tetrad. The wave vector $\varepsilon^\mu(p,\lambda)$
enables one to construct the spin-1 projection operator
\begin{eqnarray}
\varepsilon^\mu(p,\lambda)
\varepsilon^{*\mu} (p,\lambda^\prime)
=\frac{1}{3}\left[ I^{\mu\nu}
 -\frac{3}{2m}i\epsilon^{\mu\nu\rho\tau}p_\rho n_{i\tau}
  (S^i)-\frac{3}{2}n^\mu_i n^\nu_j
 (S^{ij})\right]_{\lambda^\prime \lambda},
\end{eqnarray}
where $I^{\mu\nu}=-g^{\mu\nu} + p^\mu p^\nu / m^2$, the
$S^i$ is the standard spin-1 angular momentum matrix, and
traceless, symmetric matrix $S^{ij}$ is
\begin{eqnarray}
(S^{ij})_{\lambda \lambda^\prime}
=(S^i S^j + S^j S^i -\frac{4}{3}\delta^{ij} I)
_{\lambda \lambda^\prime}.
\end{eqnarray}

When a spin-1 particle is produced, the general form of the
transition amplitude reads
\begin{eqnarray}
{\cal M} = T_\mu \varepsilon^{*\mu}(p,\lambda),
\end{eqnarray}
and the physical properties of the outgoing spin-1 particle
are determined by a vector field $\phi^\mu$ defined as
\begin{eqnarray}
\phi^\mu
&=&\sum_{\lambda} T_\alpha \varepsilon^{*\alpha}(p,\lambda)
                         \varepsilon^{\mu}(p,\lambda)
\nonumber\\
&=&\sum_{\lambda}{\cal M}(\lambda)\varepsilon^\mu(p,\lambda).
\label{pimu}
\end{eqnarray}
The covariant density matrix is then obtained as
\begin{eqnarray}
\rho^{\mu\nu}&=&-\frac{<\phi^\mu \phi^{*\nu}>}
{g_{\alpha\beta}< \phi^\alpha \phi^{*\beta}> }\nonumber\\
&=&\sum_\lambda \sum_{\lambda^\prime}
   \varepsilon^{\mu}(p,\lambda)\rho_{\lambda\lambda^\prime}
   \varepsilon^{*\nu}(p,\lambda^\prime),
   \label{rho1}
\end{eqnarray}
where the bracket $< \ \ >$ denotes the ensemble averaged
value, and $\rho_{\lambda\lambda^\prime}$ is another type
of the polarization density matrix obtained by folding
$\rho^{\mu\nu}$ with the projection operator
$\varepsilon^*_\mu(p,\lambda) \varepsilon_\nu
(p,\lambda^\prime)$ as
\begin{eqnarray}
\rho_{\lambda\lambda^\prime}&=&\frac{{\cal M} (\lambda)
{\cal M}_(\lambda^\prime)}{Tr(|{\cal M}|^2)}
\nonumber\\
&=&\sum_\mu \sum_{\nu}
   \varepsilon_{\mu}^* (p,\lambda)\rho^{\mu\nu}
   \varepsilon_{\nu}(p,\lambda^\prime).
\label{77}
\end{eqnarray}
The density matrix $\rho^{\mu\nu}$ and
$\rho_{\lambda\lambda^\prime}$ can be decomposed into three
parts;
\begin{eqnarray}
\rho^{\mu\nu}&=&\frac{1}{3} I^{\mu\nu}
-\frac{i}{2m}\epsilon^{\mu\nu\lambda\tau}p_{\lambda}P_\tau
      -\frac{1}{2}Q^{\mu\nu}, \label{8a}\\
\rho_{\lambda \lambda^\prime}
&=&\frac{1}{3}\delta_{\lambda \lambda^\prime}
-\frac{1}{2}P_\mu n_i^\mu (S^i)_{\lambda \lambda^\prime}
+\frac{1}{4}Q_{\mu\nu} n_i^\mu n_j^\nu
(S^{ij})_{\lambda \lambda^\prime},
\label{8b}
\end{eqnarray}
with the polarization variables $P^\mu$ and $Q^{\mu\nu}$ called
the covariant vector and tensor polarization, respectively.
Conversely, for a given density matrix $\rho^{\mu\nu}$, the
$P^\mu$ and $Q^{\mu\nu}$ are given by
\begin{eqnarray}
P^\mu&=&\frac{i}{m}\epsilon^{\mu\alpha\beta\gamma}
	\rho_{\alpha\beta}p_\gamma,\nonumber\\
Q^{\mu\nu}
    &=&\frac{2}{3}I^{\mu\nu}-(\rho^{\mu\nu}+\rho^{\nu\mu}).
\end{eqnarray}

The polarization effect of the initial photon beam
as in the $\gamma e$ reactions can be obtained
easily in the c.m. frame using the Coulomb gauge.
But, sometimes it is convenient to choose the polarization
vector of the photon in a covariant form as
\begin{eqnarray}
\varepsilon^\mu(\lambda)=-\frac{1}{\sqrt{2}}
(\lambda n_1^\mu + i n_2^\mu).
\label{229}
\end{eqnarray}
where $\lambda$ is $\pm 1$ for right/left handed
circular polarization and $n^\mu_1, n^\mu_2$ are
different from the $n^\mu_i$ which are
defined in massive spin-1 particle case in
Eq.~(\ref{Wvec}).
A natural cartesian basis for a polarization vector
$\varepsilon^\mu(\lambda)$ with a momentum $k_1$ can be given
in terms of two arbitrary four momenta $p_1$ and $p_2$
satisfiying the
constraints $p_1^2=p_2^2=0$.
For simplicity, we use $p_1$ and $p_2$
for the momenta of the incident electron and final
fermion in the $\gamma e$ reaction.
Then we can choose the basis consisting of two orthonomal
four-vector $n_1$ and $n_2$ such that \cite{Calkul}
\begin{eqnarray}
n_1^\mu &=& N[(k_1\cdot p_2)p_1^\mu -(k_1\cdot p_1)p_2^\mu],
\label{231}\\
n_2^\mu &=& N\epsilon^{\mu\alpha\beta\gamma}
	    k_{1\alpha} p_{1\beta} p_{2\gamma},\label{232}
\end{eqnarray}
where $N$ is a normalization factor given by
\begin{eqnarray}
N={[2(k_1\cdot p_1)(k_1\cdot p_2)(p_1\cdot p_2)]}
   ^{-\frac{1}{2}}
.\label{233}
\end{eqnarray}
In the covariant density matrix formalism, the photon
polarization operator $\varepsilon^\mu (\lambda)
\varepsilon^{*\nu} (\lambda^\prime)$ for an incident
photon beam is replaced by its photon covariant density
matrix
\begin{eqnarray}
\rho^{\mu\nu}
=&&\frac{1}{2}[ (n_1^\mu n_1^\nu +n_2^\mu n_2^\nu)
+\xi_1 (n_1^\mu n_2^\nu +n_2^\mu n_1^\nu)
\nonumber\\
&&-i\xi_2 (n_1^\mu n_2^\nu -n_2^\mu n_1^\nu)
+\xi_3 (n_1^\mu n_1^\nu -n_2^\mu n_2^\nu)],
\label{star}
\end{eqnarray}
where the $\xi_i$ are Stokes parameters of the
photon beam.

In the process $\gamma e \rightarrow Z e$, the wave vector
$\phi^\mu$ of the $Z^0$ vector boson can be described in
the SM as
\begin{eqnarray}
&&\phi^\mu_Z=
\frac{e}{2}g_Z
   \bar{u}(p_2,s_2)
\left[
 \frac{\gamma_\alpha(\not\!{k_1}\not\!{\varepsilon}
 -2p_1\cdot \varepsilon)} {k_1\cdot p_1}
 - \frac{(2p_2\cdot \varepsilon
       -\not\!{\varepsilon} \not\!{k_1})\gamma_\alpha}
    {k_1\cdot p_2} \right]
\nonumber\\
 &&\hskip 2cm \times \left[
   \epsilon_L(1-\gamma_5) +\epsilon_R(1+\gamma_5)
     \right] u(p_1,s_1)
      \sum_{\lambda}
      \varepsilon_Z^{*\alpha}(k_2, \lambda)
      \varepsilon_Z^{\mu}(k_2, \lambda),
\label{rho2}
\end{eqnarray}
where $\varepsilon^\mu$ implies the wave vector of the photon
, i.e., $\varepsilon^\mu_\gamma$.

Then the density matrix $\rho^{\mu\nu}$ of $Z^0$ can be
obtained from Eqs.~(\ref{rho1}) and (\ref{rho2}) explicitly.
In particular, one obtain the denominator of Eq.~(\ref{rho1})
as
\begin{eqnarray}
&&-g^{\mu\nu}<\phi_Z^{\mu} \phi_Z^{*\nu}>
 =\frac{e^2 g_Z^2 }{2s^2 u^2}A_0,
 \label{gmn}
 \\
&&A_0=
(|\epsilon_R|^2 +|\epsilon_L|^2)[\varepsilon \cdot
         \varepsilon^* (s^2+u^2)su
	 -4m_Z^2 \varepsilon \cdot A
	     \varepsilon^* \cdot A]
\nonumber\\
&&\hskip 1cm
 +2 i(|\epsilon_R|^2 -|\epsilon_L|^2)s(s-u)(t+m_Z^2)
           <\varepsilon \varepsilon^* k_1 p_2>
	   \label{deno},
\end{eqnarray}
where $s,t$, and $u$ are the usual Mandelstam variables,
and $A^\mu$ and
$<\varepsilon \varepsilon^* k_1 p_2>$ are defined as
\begin{eqnarray}
A^\mu&=&up_1^\mu + s p_2^\mu, \nonumber\\
<\varepsilon \varepsilon^* k_1 p_2>
&=&\epsilon_{\mu\nu\lambda\tau}
\varepsilon^\mu \varepsilon^{*\nu}
	   k_1^\lambda p_2^\tau.
\end{eqnarray}
Then the differential cross section of the $\gamma e
\rightarrow Z e$ can be obtained from $A_0$ as
\begin{eqnarray}
&&\left[\frac{d\sigma}{dt}\right]_{\gamma e \rightarrow Z e}
 =\frac{e^2 g_Z^2}{64\pi s^4 u^2}A_0,
 \nonumber\\
&&\hskip 2.0cm =\frac{e^2 g_Z^2}{64\pi s^4 u^2}
\left\{(|\epsilon_R|^2 +|\epsilon_L|^2)[\varepsilon \cdot
         \varepsilon^* (s^2+u^2)su
	 -4m_Z^2 \varepsilon \cdot A
	     \varepsilon^* \cdot A]\right.
\nonumber\\
&&\hskip 4.2cm \left.
 +2 i(|\epsilon_R|^2 -|\epsilon_L|^2)s(s-u)(t+m_Z^2)
           <\varepsilon \varepsilon^* k_1 p_2>\right\}
	   \label{dsz}
\end{eqnarray}
If the incident electron beam is polarized, then
$|\epsilon_R|^2$ and $|\epsilon_L|^2$ in Eq.~(\ref{deno})
should be replaced by $(1+\hat{s_1}\cdot \hat{p_1})
|\epsilon_R|^2/2$ and $(1-\hat{s_1}\cdot \hat{p_1})
|\epsilon_L|^2/2$, respectively.
Also, Eq.~(\ref{deno}) can be used in the process $e^+ e^-
\rightarrow Z \gamma$, if
$\varepsilon^\mu, k_1$ and $p_2$ are replaced by
$\varepsilon^{*\mu}, -k_1$ and $-p_2$ so that $s$ and
$t$ are interchanged, and the explicit form in the c.m.
frame was considered in Ref. \cite{JC}.

It is found that in the SM Eqs.~(\ref{gmn}) and
(\ref{deno}) can be used for the $\gamma e \rightarrow
W \nu$ process after $\epsilon_R$ is neglected and also
$m_Z$ and $g_Z$ are replaced by $m_W$ and $g_W$.
Therefore, one can obtain the differential cross section
immediately as following,
\begin{eqnarray}
&&\left[
\frac{d\sigma}{dt}\right]_{\gamma e \rightarrow W \nu}
 =\frac{e^2 g_W^2}{64\pi s^4 (s+u)^2}
\left[ \varepsilon \cdot
         \varepsilon^* (s^2+u^2)su
	 -4m_W^2 \varepsilon \cdot A
	     \varepsilon^* \cdot A
	     \right.
\nonumber\\
&&\hskip 5.5cm \left.	-2 is(s-u)(t+m_W^2)
           <\varepsilon \varepsilon^* k_1 p_2>\right].
\label{dsdt}
\end{eqnarray}
In particular, if the c.m. frame of $\gamma$ $e$ and the
Coulomb gauge are considered \cite{JC}, one obtains
\begin{eqnarray}
 \varepsilon \cdot \varepsilon^*&=&-1,
\nonumber\\
 \varepsilon \cdot A \varepsilon^* \cdot A
 &=&\frac{1}{2}stu(1+\xi_3 \cos 2\phi -\xi_1\sin 2\phi),
 \nonumber\\
 <\varepsilon \varepsilon^* k_1 p_2>
   &=&-\frac{i}{2}u \xi_2,
 \label{epsilon}
\end{eqnarray}
where the azimuthal angle $\phi$ which specifies the
linear polarization of the photon through the Stokes
parameters $\xi_i (i=1,3)$ can be chosen to be $0$
without the loss of generality \cite{JC,JCh}.
The differential cross section becomes
\begin{eqnarray}
&&\left[
\frac{d\sigma}{dt}\right]_{\gamma e \rightarrow W \nu}
 =-\frac{e^2 g_W^2 u}{64\pi s^3 (s+u)^2}
\left[ (s^2+u^2)
	 +2m^2 t(1+\xi_3 \cos 2\phi -\xi_1\sin 2\phi),
	     \right.
\nonumber\\
&&\hskip 5.8cm \left. +\xi_2 (s-u)(t+m^2)
        \right]\label{dsw}
\end{eqnarray}
The result of Eq.~(\ref{dsdt}) is of a covariant form
and it can be obtained by using the covariant photon
polarization, Eq.~(\ref{star}), but then one obtain
the $\phi=0$ case.
It is noted that, if the incoming photon beam is
unpolarized, the $\xi_i$ become zero and the result
is the same as that in Ref. \cite{LF}.
Also the relation in the square bracket of
Eq.~(\ref{dsw}) is the same as that of Eq.~(\ref{dsz})
in the process $\gamma e \rightarrow Z e$ except
$\epsilon_R=0$, and it
differs from Eq. (4.3) of Ref. \cite{FMR2} which
must be corrected.
On the other hand, one can obtain the same result
of differential cross section for the process
$\gamma e \rightarrow Z e$ as that
in Ref. \cite{FMR2}
in the center of mass frame of $\gamma$ and $e$
from Eqs.~(\ref{dsz}) and (\ref{epsilon})
when the incident photon beam is polarized.

{}From the explicit form of $\rho^{\mu\nu}$, one obtains
explicit values of polarization vector $P^\mu$ and
polarization tensor
$Q^{\mu\nu}$ in the process $\gamma e \rightarrow Z e$
as
\begin{eqnarray}
&&P^\mu=\frac{2}{m_Z}
   \left[i(|\epsilon_R|^2 +|\epsilon_L|^2)R_1^\mu
             + (|\epsilon_R|^2 -|\epsilon_L|^2)R_2^\mu
		\right]/A_0, \nonumber\\
&&Q^{\mu\nu}= -\frac{1}{3}I^{\mu\nu}
    - 2\left[(|\epsilon_R|^2 +|\epsilon_L|^2)R_1^{\mu\nu}
         + i(|\epsilon_R|^2 -|\epsilon_L|^2)R_2^{\mu\nu}
		\right]/A_0,
		\label{pqmu}
\end{eqnarray}
where $R_1^\mu, R_2^\mu, R_1^{\mu\nu}$ and $R_2^{\mu\nu}$,
are defined as
\begin{eqnarray}
&&R_1^\mu=-s
     <\varepsilon  \varepsilon^* k_1 p_2>
     \left[2m_Z^2(tk_1^\mu-sp_1^\mu+up_2^\mu)
     +(s^2+u^2)k_2^\mu\right],
\nonumber\\
&&R_2^\mu=m_Z^2\left[su( \varepsilon ^\mu
                   \varepsilon^* \cdot A
                   +\varepsilon^{*\mu}
                   \varepsilon \cdot A)
   -su(s-u) \varepsilon \cdot \varepsilon^* K^\mu
   \right.
\nonumber\\
&&\left.\hskip 2cm
    -2(p_1^\mu+p_2^\mu)\varepsilon \cdot A
                   \varepsilon^* \cdot A
   -2k_1^\mu(u^2 \varepsilon \cdot p_1
                   \varepsilon^* \cdot p_1
          -s^2 \varepsilon \cdot p_2
               \varepsilon^* \cdot p_2)\right],
\nonumber\\
&&R_1^{\mu\nu}=su\varepsilon \cdot \varepsilon^*
    \left[su I^{\mu\nu}-2m^2 K^\mu K^\nu\right]
   +\left[(su\varepsilon ^\mu
     -	\varepsilon _\alpha B^{\alpha\mu})
    (su	\varepsilon ^{*\nu}
     -	\varepsilon ^* _\alpha B^{\alpha\nu})
     +(\mu \leftrightarrow \nu)\right],
\nonumber\\
&&R_2^{\mu\nu}=s
	  <\varepsilon \varepsilon^* k_1 p_2>
	  \left\{(u-s) (t+m_Z^2)
          \frac{k_2^\mu k_2^\nu}{m_Z^2}
  +2\left[m_Z^2k_1^\mu(p_1+p_2)^\nu-k_2^\mu A^\nu
        +(\mu \leftrightarrow \nu)\right]\right\},
	  \label{qrmu}
\end{eqnarray}
where $K^\mu$, and $B^{\alpha\mu}$, and
$<\mu \varepsilon \varepsilon^* k_1>$
are defined as
\begin{eqnarray}
&&K^\mu=k_1^\mu -\frac{k_1\cdot k_2}{m_Z^2}k_2^\mu
         =k_1^\mu -\frac{(s+u)}{2m_Z^2}k_2^\mu,
\nonumber\\
&&B^{\alpha\mu}=u p_1^\alpha (2p_2+k_2)^\mu
              +s p_2^\alpha (2p_1-k_2)^\mu
              = (p_1+p_2)^\mu A^\alpha
              + k_1^\mu(u p_1^\alpha-s p_2^\alpha),
\nonumber\\
&&<\mu \varepsilon \varepsilon^* k_1>=
 \epsilon^{\mu\nu\alpha\beta}
 \varepsilon_\nu \varepsilon^*_\alpha k_{1\beta}.
\end{eqnarray}
The corresponding relations of Eq.~(\ref{qrmu})
in the process $e^+ e^- \rightarrow Z \gamma$
was given in Ref. \cite{JC}, where the c.m.
frame of $e^+$ and $e^-$
and Coulomb gauge for the photon beams are
chosen, and they can be obtained from
Eqs.~(\ref{pqmu})-(\ref{qrmu}) as a special case.
Moreover, the above results can be used to obtain
for the polarization vector ${P_W}^\mu$ and polarization
tensor ${Q_W}^{\mu\nu}$ of the $W$ boson in the process
$\gamma e \rightarrow W \nu$, by putting
$\epsilon_R=0$ and by replacing $m_Z$ by $m_W$ in Eqs.~
(\ref{deno}) and (\ref{qrmu}).

As we can see in Eq.~(\ref{star}), $\xi_1$ and $\xi_3$
terms are symmetric in the indices $\mu$ and $\nu$, and
the $\xi_2$ term is antisymmetric in those.
Therefore, $R_1^\mu$ and $R_2^{\mu\nu}$ in
Eq.~(\ref{qrmu}) depend on the degree of circular
polarization of the incident photon beams, $\xi_2$,
while $R_2^\mu$ and $R_1^{\mu\nu}$ can be obtained
in terms of the linearly polarized parameters
$\xi_1$ and $\xi_3$.
Explicitly
the influence of the photon polarization in the reactions
can be seen through $A_0, R_i^\mu$, and $R_i^{\mu\nu}$
$(i=1,2)$ as following,
\begin{eqnarray}
&&A_0 =su\left\{-(|\epsilon_R|^2+|\epsilon_L|^2)
    \left[2t m_Z^2(1+\xi_3)+ s^2 +u^2 \right]
   +\xi_2(|\epsilon_R|^2-|\epsilon_L|^2) (s-u)(t+m_Z^2)
    \right\},
\nonumber\\
&&R_1^\mu= \frac{i}{2}su \xi_2
     \left[2m_Z^2(tk_1^\mu-sp_1^\mu+up_2^\mu)
     +(s^2+u^2)k_2^\mu\right],
           \nonumber\\
&&R_2^\mu=su m_Z^2\left\{ (s-u)K^\mu
        -(1+\xi_3)\left[A^\mu+t(p_1+p_2)^\mu\right]
        + 2\xi_1<\mu k_1 p_1 p_2>
          \right\} ,
          \nonumber\\
&&R_1^{\mu\nu}= \frac{su}{t}\left\{
         -t(su I^{\mu\nu}-2m_Z^2 K^\mu K^\nu)\right.
  \nonumber\\
    &&  \hskip 1cm -(1-\xi_3)
   \left[ t^2 k_1^\mu k_1^\nu +sut g^{\mu\nu}
      -tk_1^\mu(up_1^\nu-sp_2^\nu)
      -tk_1^\nu(up_1^\mu-sp_2^\mu)
      +A^\mu A^\nu\right]
\nonumber\\
 &&\hskip 1cm
  +(1+\xi_3)\left[A^\mu+t(p_1+p_2)^\mu\right]
              \left[A^\nu+t(p_1+p_2)^\nu\right]
\nonumber\\
  &&\hskip 1cm \left.  - 2\xi_1\left[<\mu k1 p1 p2>
              \left(A^\nu+t(p_1+p_2)^\nu\right)
               +(\mu \leftrightarrow \nu)\right],
 \right\}
\nonumber\\
&&R_2^{\mu\nu}=-\frac{i}{2}su \xi_2
	  \left\{(u-s) (t+m_Z^2)
            \frac{k_2^\mu k_2^\nu}{m_Z^2}
     +2\left[m_Z^2k_1^\mu(p_1+p_2)^\nu-k_2^\mu A^\nu
         +(\mu \leftrightarrow \nu)\right]\right\}.
\end{eqnarray}


\section{Decay of Vector Bosons}
\label{sec:decay}

Once the vector bosons like $Z^0$'s or $W$'s are produced,
they have the polarization vector $P^\mu$ and polarization
tensor $Q^{\mu\nu}$ and they decay into lighter particles
immediately. The density matrix for the vector mesons can
be used in this case.
Since the $Z^0$ decay has been considered previously
\cite{JC,Calkul}, consider here the case that a
$W$ decays into a lepton and a neutrino.

In the SM, the transition amplitude for the decay
$W \rightarrow l \bar{\nu_l}$ becomes
\begin{eqnarray}
{\cal M} = g_W  \varepsilon^\mu_W
    \bar{u}(p_1^\prime)\gamma_\mu(1-\gamma_5)v(p_2^\prime),
\label{decm}
\end{eqnarray}
where $p_1^\prime$ and $p_2^\prime$ are the momenta of
a lepton and a neutrino, respectively, and
$\varepsilon^\mu_W$ is the Proca vector of the $W$ before
its decay.
The full amplitude for the production of $W$ followed by
its subsequent decay can be obtained if
$\varepsilon^\mu_W$ in Eq.~(\ref{decm}) is replaced by
$\phi^\mu_W$ in Eq.~(\ref{pimu}) and multiplied by the $W$
propagator in the Wigner-Breit form.

If the outgoing lepton is unpolarized, the absolute
square of the amplitude, after replacing
$\varepsilon^\mu_W \varepsilon^{*\nu}_W$ by
$\rho^{\mu\nu}$, becomes
\begin{eqnarray}
\sum_f |{\cal M}|^2
= \frac{4g_W^2}{m_l m_\nu}\rho^{\mu\nu}
   \left[p_{1\mu}^\prime p_{2\nu}^\prime
  +p_{1\nu}^\prime p_{2\mu}^\prime
  -p_1^\prime\cdot p_2^\prime g^{\mu\nu}
  +i \epsilon_{\mu\nu\lambda\tau}
        p_1^{\prime\lambda} p_2^{\prime\tau}\right].
\end{eqnarray}

{}From the explicit expression of $\rho^{\mu\nu}$ in terms
of $P^\mu$ and $Q^{\mu\nu}$ as in Eq.~(\ref{8a}), one
obtains the angular distribution for the outgoing lepton
as
\begin{eqnarray}
\frac{d\Gamma}{d \Omega}
= \frac{m_W g_W^2}{48\pi^2}
   \left[1+ \frac{3}{m_W} P\cdot p_1^\prime
          + \frac{3}{m_W^2} Q^{\mu\nu}
  p_{1\mu}^\prime p_{1\nu}^\prime
        \right].
\label{dGO}
\end{eqnarray}
Since the effect of photon polarization is contained in
$P^\mu$ and $Q^{\mu\nu}$, the asymmetry due to the
photon polarization can be obtained from Eq.~(\ref{dGO}).

If $\rho^{\mu\nu}$ is replaced by Eq.~(\ref{rho1}), one
obtains
\begin{eqnarray}
\sum |{\cal M}|^2
= \frac{4g_W^2}{m_l m_\nu}
  \sum_\lambda \sum_{\lambda^\prime}
  \rho_{\lambda\lambda^\prime}
   \left[p_1^\prime\cdot p_2^\prime
    \delta_{\lambda\lambda^\prime}
  -2\varepsilon(\lambda)\cdot p_1^\prime
  \varepsilon^*(\lambda^\prime)\cdot p_1^\prime
  +i<\varepsilon(\lambda) \varepsilon^*(\lambda^\prime)
      p_1^\prime p_2^\prime>\right]
\label{m36}
\end{eqnarray}
In particular, when the $W$ decays in its rest frame,
the $n_i^\mu (i=1,2,3)$ becomes $\hat{\theta}, \hat{\phi}$,
and $\hat{s}$ which can be chosen as unit vectors along
the $x,y$, and $z$ direction, respectively.
Using the explicit form of $\varepsilon^\mu (\lambda)$
in Eq.~(\ref{m36}) given by Eq.~(\ref{Wvec}),
one obtains the angular distribution of the outgoing lepton
in terms of density matrix elements
$\rho_{\lambda\lambda^\prime}$ as
\begin{eqnarray}
&&I(\theta,\phi)
=\frac{3}{16\pi}\left[
1+\cos^2\theta+\rho_{00}(1-3\cos^2\theta)
-2\cos \theta (\rho_{11}-\rho_{-1-1})
+2\sin^{2}\theta \ \
Re\left\{\rho_{1-1} e^{2i\phi}\right\}
\right.
\nonumber\\
&&\left. \hskip 2.5cm -2\sqrt{2}\sin \theta \ \
Re\left\{(\rho_{10} +\rho_{0-1}) e^{i\phi}\right\}
 +\sqrt{2}\sin 2\theta \ \
Re\left\{(\rho_{10} -\rho_{0-1}) e^{i\phi}\right\}
\right]
\end{eqnarray}

The density matrix elements can be obtained experimentally
by investigating the angular dependence of the outgoing
lepton beam.
On the other hand, $\rho_{\lambda\lambda^\prime}$
can be obtained from Eqs.~(\ref{77}) and (\ref{8b}) and,
in particular, in the $W$ rest frame it becomes
\begin{eqnarray}
\rho_{\lambda\lambda^\prime}=
\left[\frac{1}{3}I+\frac{1}{2} P^i S^{i}
+\frac{1}{4} Q^{ij}S^{ij}
\right]_{\lambda\lambda^\prime}
\end{eqnarray}
Therefore, it depends on the polarization vector $\vec{P}$
and polarization tensor $Q^{ij}$ again, and explicit values
$\rho_{\lambda \lambda^\prime}$ in Eq.~(\ref{dsdt}) depends
on the axis one chooses,
\begin{eqnarray}
&&\rho_{00}=\frac{1}{3}-\frac{1}{2}Q^{33},\nonumber\\
&&\rho_{11}-\rho_{-1-1}=P^{3},\nonumber\\
&&\rho_{10}+\rho_{0-1}=\frac{1}{\sqrt{2}}
       (P^1-iP^2),
\nonumber\\
&&\rho_{10}-\rho_{0-1}=
     \frac{1}{2\sqrt{2}}(Q^{13}-iQ^{23}),\nonumber\\
&&\rho_{1-1}=\frac{1}{4}
     (Q^{11}-Q^{22}-iQ^{12}),
\end{eqnarray}
where explicit forms of $\vec{P}$ and $Q^{ij}$ in the $W$
rest frame are given as following
\begin{eqnarray}
&&P^i=-\frac{2m_W[(s+t)k_1^i + (t+m_W^2)p_1^i]}
            {2m_W^2 t+s^2+u^2} \nonumber\\
&&Q^{ij}=-\frac{1}{3} \delta ^{ij}
         +\frac{2m_W^2 (k_1^i k_1^j +2 p_1^i p_1^j)
           + (t+m_W^2)(p_1^i k_1^j+p_1^jk_1^i)}
            {2m_W^2 t+s^2+u^2}.
\end{eqnarray}

Usually, the $x-z$ axis is chosen in the production plane.

\section{Conclusion and Discussion}
\label{sec:con}


The polarization vector $P^\mu$ and polarization tensor
$Q^{\mu\nu}$ in the $\gamma e \rightarrow Z e$ process
are obtained explicitly by means of the covariant
density matrix formalism. Using the crossing symmetry,
these results can be used to obtain the polarization
of $Z^0$ beam produced by the $e^+ e^- \rightarrow Z
\gamma$ process.
It is shown that in the SM, the polarization of the $W$
boson polarization in the $\gamma e
\rightarrow W \nu$ process can be obtained from that
of $Z^0$ in the $\gamma e \rightarrow Z e$ process
just by putting $\epsilon_R=0$ and by replacing
$m_Z$ and $g_Z$
by $m_W$ and $-(k_1\cdot p_2/k_1\cdot k_2) g_W$
, respectively.

Since the results are of the covariant form, they can be
used in any frame, e.g., in the c.m. frame or in the
rest frame of the vector bosons. The usual results
discussed in the c.m. frame using the helicity formalism
can be reproduced from the covariant results.

Some possible deviations from the SM have been
considered by several authors through the anomalous
gauge-boson-couplings, in the CP conserving cases
\cite{MG,LF,EY,MR}
as well as CP violating cases \cite{DC,SY1},
by introducing additional terms.
Such additional terms change $A_0$, $P^\mu$, and
$Q^{\mu\nu}$. Therefore, the universality considered
in the SM would not be held in general, and the
underlying models for such changes can be investigated
through the polarization effects of
vector bosons as well as those of fermion and photon
involved in the reactions.

\section*{Acknowledgements}

The authors would like to thank S. Y. Choi for useful
comments.
The work was supported in part by the Korea Science
and Engineering Foundation through the SRC program
and in part by the Korean Ministry of Education.
J. S. S would like to thank Kyung Hee University
for financial assistance.


\newpage
\section*{Figure Captions}
\begin{enumerate}
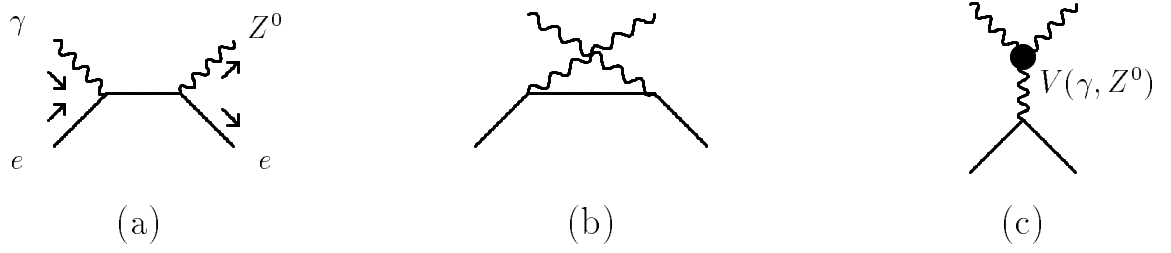
\item[{\bf Fig.~1}]
     Feynman diagrams for the process
     $\gamma e \rightarrow Z^0 e$ in
     the SM $(a)$ and $(b)$, and the
     anomalous contribution $(c)$.
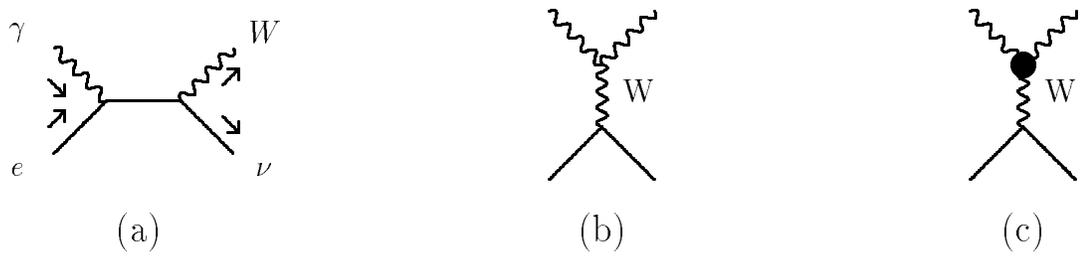
\item[{\bf Fig.~2}]
     Feynman diagrams for the process
     $\gamma e \rightarrow W \nu$ in
     the SM $(a)$ and $(b)$, and the
     anomalous contribution $(c)$.
\item[{\bf Fig.~3}]
     Feynman diagrams for the process
     $e^+ e^- \rightarrow Z^0 \gamma$ in
     the SM $(a)$ and $(b)$, and the
     anomalous contribution $(c)$.
\end{enumerate}

\centerline{\epsfig{file=grim.eps}}
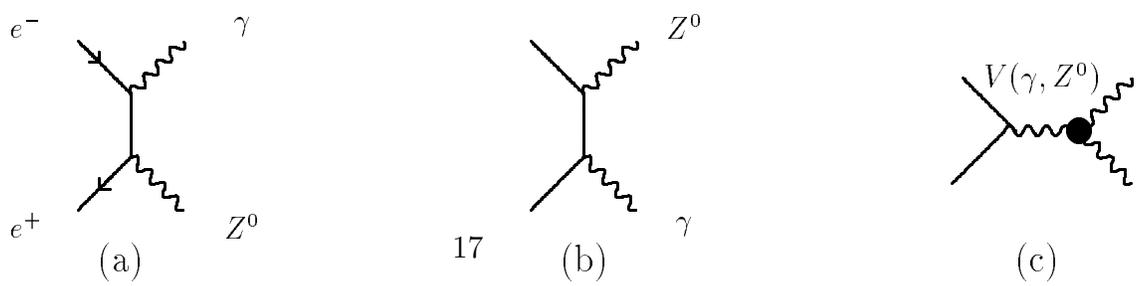
\end{document}